\documentclass[doublespacing]{elsart}

\usepackage{graphicx}
\usepackage{natbib}
\usepackage{lineno}
\journal{Ecological Complexity}

\begin{document}

\begin{frontmatter}
\title{Sea urchin feeding fronts}

\author{Edward R. Abraham\thanksref{now}}
\address{National Institute of Water and Atmospheric Research (NIWA),
P.O. Box 14-901, Kilbirnie, Wellington, New Zealand}
\ead{edward@dragonfly.co.nz}

\thanks[now]{Present address: Dragonfly, 10 Milne Terrace, Island Bay, Wellington, New Zealand (www.dragonfly.co.nz)}

\begin{abstract}
Sea urchin feeding fronts are a striking example of spatial pattern formation in an ecological system. If it is assumed that urchins are asocial, and that they move randomly, then the formation of these dense fronts is an apparent paradox. The key lies in observations that urchins move further in areas where their algal food is less plentiful. This naturally leads to the accumulation of urchins in areas with abundant algae. If urchin movement is represented as a random walk, with a step size that depends on algal concentration, then their movement may be described by a Fokker-Planck diffusion equation. For certain combinations of algal growth and urchin grazing, travelling wave solutions are obtained. Two dimensional simulations of urchin algal dynamics show that an initially uniformly distributed urchin population, grazing on an alga with a smoothly varying density, may form a propagating front separating two sharply delineated regions. On one side of the front algal density is uniformly low, and on the other side of the front algal density is uniformly high. Bounds on when stable fronts will form are obtained in terms of urchin density and grazing, and algal growth.
\end{abstract}
\end{frontmatter}

\linenumbers
\modulolinenumbers[2]

\section{Introduction}
\label{sec:intro}

Dense, linear aggregations of sea-urchins are sometimes seen. These
features, known as feeding-fronts, generally occur at the boundary
between grazed and ungrazed habitat \citep{Dean84, Scheibling99, Alcoverro02, Gagnon04}. The fronts propagate slowly
towards the ungrazed region. Because of the high urchin densities,
they are often destructive. A striking example was an
aggregation of the urchin \emph{Lytechinus variegatus},
observed invading sea-grass habitat in Florida Bay \citep{Macia99}.
The aggregation was estimated to be 2 - 3 m wide and 4 km long, with a
density of order 100 urchins m$^{-2}$. It is reported to have moved at a rate of up to 6 m
day$^{-1}$, reducing above-ground seagrass to less than 2\% of its
initial biomass. Although it became more diffuse with time, the front
remained as a coherent feature for at least 10 months. Similar features have been seen in other benthic invertebrates. Linear
aggregations of starfish have been recorded invading extensive mussel
beds \citep{Dare82}, and traveling fronts of strombid conch have also
been observed in the Caribbean \citep{Stoner89, Stoner94} and in
Australia (A.  MacDiarmid, pers.~comm.). Because of the strong
influence of such aggregations on the benthic habitat, it is
interesting to question how they are formed and maintained.

Herds, flocks, schools, and swarms are all aggregations of social
animals. The aggregation is caused by the interaction between the
individuals, which attracts them together at large distances
\citep{Okubo80}. For animals such as sea-urchins there is little
evidence that they are social. In uniform habitat their clumping is
mild \citep{Andrew86, Hagen95}. Experiment suggests that urchins will
aggregate in the presence of food \citep{Vadas86}, but there is no
evidence for a strong social interaction. Moreover, studies of urchin
movement have found that while they may exhibit a chemosensory
response to algae, they do not show any directed movement towards
it \citep{Andrew86}. A recent flume tank study shows that the urchin \emph{Lytechinus variegatus} can move in a directed manner towards a food source under some flow conditions \citep{Pisut02}. This may explain how urchins locate their food at short distances. Both the flow and the chemical signals are likely to be more complex in the urchins' natural environment. In field studies the direction of urchin movement is usually found to be either random or weakly directional \citep{Duggan01, Dumont06, Lauzonguay06}. The question then is how to explain the formation
of intense aggregations in an asocial animal, which appears not to be
able to move in a directed manner.

\begin{figure}
	\begin{center}
\includegraphics*[width=7cm]{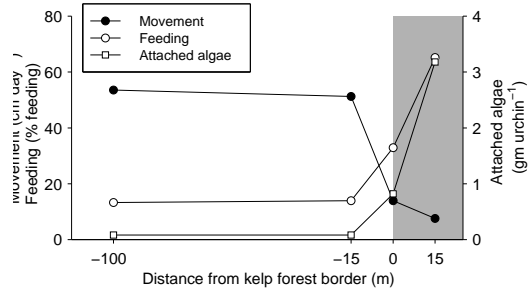}
\end{center}
\caption{Movement of red sea urchins,
  \emph{Strongylocentrotus franciscanus}, near the boundary of a kelp
  forest at Santa Cruz Point \citep[redrawn from ][]{Mattison77}. The
  figure shows the average rate of urchin movement, measured over a 24
  hour period, at four locations. For comparison, the percentage of
  urchins which were observed to be feeding, and the weight of algae
  attached to the urchins' oral surface, are also shown. Within the
  kelp forest (shaded), feeding is high and movement rates are low.}
\label{fig:mattison}
\end{figure}

A recurrent observation is that there is an inverse relation between urchin movement and macrophyte density
\citep{Mattison77, Andrew86, Dance87, Dumont06}. A study by \citet{Mattison77} of
red sea-urchins (\emph{Strongylocentrotus franciscanus}) near Santa
Cruz found that urchins within a kelp forest moved by 7.5 cm
day$^{-1}$, whereas outside it the movement rate increased to over 50
cm day$^{-1}$ (Fig.~\ref{fig:mattison}). The reasons for the difference in movement rates between habitats is not clear. Some studies find that movement rate is more for starved urchins \citep{Dix70, Hart90}, whereas others find either no effect \citep{Dumont06} or the opposite relation \citep{Klinger85}. It has also been shown, by using physical models of large algae, that the movement of foliose algae by the water may restrict urchin movement \citep{Konar03}. In this paper, the consequences of differential motility in different habitats will be explored, whatever its cause. Four simple assumptions are made about sea urchin movement:
\begin{enumerate}
\item Sea urchins are asocial, with the movements of individual
  urchins being independent
\item The direction of sea urchin movement is random (over a suitable
  time period, which we take to be 24 hours)
\item The sea-urchin
  movement rate decreases as the macrophyte density increases
\item The distance moved in a 24 hour period is related to the seaweed density at the beginning of the time-period. 
\end{enumerate}

The consequences of these assumptions are explored, using both
analytical techniques and direct simulation. It might seem to be intuitively reasonable that if the urchins are randomly moving then they will disperse, and it will be impossible for them to accumulate into an organised structure like a feeding front. In this paper it is shown that under
certain circumstances, and with a suitable representation of
macrophyte growth and urchin grazing, the assumptions about urchin movement may lead to persistent urchin feeding fronts.

There are other features of urchin movement which are not accounted for by this model. A recent study \citep{Lauzonguay06} of sea urchin movement, which followed the movements of individual urchins using video techniques, showed that the distance moved decreased with increasing urchin density. This effect is not included in the present study. Other authors have concluded that the urchin response to predators may mediate the formation of feeding fronts \citep{Bernstein81}. The model we discuss is a minimal model. The complexities of differential feeding on multi species algal assemblages \citep{Gagnon04, Wright05}, size dependent urchin movement \citep{Dumont04, Dumont06}, seasonal variations in movement rate \citep{Konar01, Dumont04}, relation between behaviour and the supply of drift algae \citep{Dayton84}, interactions between movement and the substrate \citep{Laur86}, or between water movement and urchin movement \citep{Kawamata98} are not included. All demographic processes such as urchin growth, recruitment and mortality have also been ignored. If sufficient data were available these processes could be represented. However, while their inclusion would lead to a more realistic model of a specific system, the purpose of this paper is to explore the consequences of a single urchin behaviour.

\section{Urchin movement and the Fokker-Planck equation}
\label{sec:movement}

\begin{figure}
\label{fig:errorfunc}
\centering
\includegraphics[width=7cm]{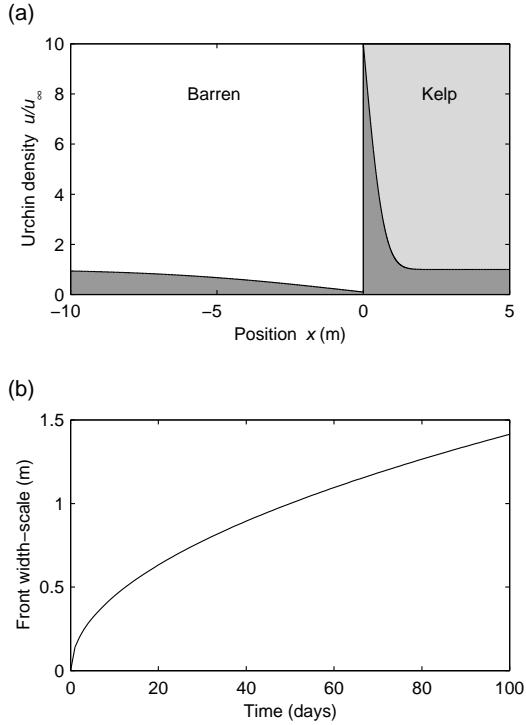}
\caption{Solution to the Fokker-Planck equation,  eq.~(\ref{eq:fp}),
  describing the dispersal of an initially uniform population of
  sea-urchins in response to a step-change in macrophyte density. The
  right-hand side, $x>0$, is kelp forest with the urchin movement being
  $\lambda_+ = 0.1$ m and the left-hand side, $x<0$, is barren with
  $\lambda_- = 1   $ m (here $\Delta t = 1$ day).  These values are
  chosen to be comparable with \citet{Mattison77}. (a) The urchin
  distribution after 30 days, with the initial population having a
  value $u/u_\infty = 1$. There is a net movement of urchins from the
  barren region to the kelp-forest, with a sharp peak appearing at the
  kelp boundary. (b) The width of the peak, $2 \sqrt{ D_+ t}$,
  increases very slowly. Even after 100 days it is less than two  meters
  wide.  The maximum urchin density is constant with time, at 10 times
  the initial population.}
\end{figure}

The four assumptions above may be used to formalize sea-urchin movement as a random walk. If $x_i(t)$ is the
position of urchin $i$ at time $t$, then its position a time $\Delta
t$ later may be represented as
\begin{equation}
  \label{eq:randomwalk}
  x_i(t+\Delta t) = x_i(t) + \eta(t)\lambda(s(x_i(t))) ,
\end{equation}
where $\eta(t)$ is a dimensionless random variable with a zero mean and a unit
variance, and $\lambda(s(x))$ (dimensions [$x$]) is a characteristic step-size which is a function of the macrophyte density,
$s$. 

If the movement of individual sea urchins satisfies eq.~(\ref{eq:randomwalk}), then the dispersal of the population may be approximated by the continuous
Fokker-Planck equation \citep{Turchin98},
\begin{equation}
  \label{eq:fp}
  {\partial u \over \partial t} = {\partial^2 \over \partial x^2}(D u) ,
\end{equation}
where $u(x, t)$ is the urchin density and the motility $D(s)$ (dimensions [$x^2$ $t^{-1}$]) is
related to the random-walk parameters by
\begin{equation}
  \label{eq:rwp}
  D(s)  = {\lambda(s)^2 \over 2 \Delta t}.
\end{equation}

The long term behavior of the population $u$ is well-known. If the total number of sea-urchins is constant with time, then the steady
state solution to eq.~(\ref{eq:fp}) is
\begin{equation}
  \label{eq:steadystate}
  u(x,t) = c/D(s),
\end{equation}
where $c$ is a constant. At equilibrium, the population density will
be inversely related to the motility. The sea-urchins will accumulate in areas where the seaweed concentration is higher, and so the individual urchins are moving more slowly. The aggregation of randomly
walking foragers in regions with higher food density, is known variously as preytaxis
\citep{Kareiva87}, orthokinesis \citep{Okubo80}, or phagokinesis
\citep{Andrew86}. An experimental study of ladybugs feeding on an
inhomogeneous aphid population showed that, in this case,
eq.~(\ref{eq:steadystate}) provided a good description of the data \citep{Turchin98}. The random walk formalism is similar to (although simpler than) that used to understand the formation of traveling bands of bacteria through chemotaxis \citep{Keller71b}.

While it has been observed that urchin movement is higher when the
algal density is lower, little is known about the functional form of
$\lambda(s)$. In the absence of any data, we will simply assume that
there is a threshold algal density, $s_c$, at which the rate of urchin
movement changes from a minimum to a maximum value,
\begin{equation}
  \label{eq:Dsc}
  D = \left\{ 
    \begin{array}{cl}
    D_-, & s<s_c \\
    D_+, & s\geq s_c \\
    \end{array}
    \right.,
\end{equation}
where $D_- > D_+ >0$. Within this model, the urchins have only two behaviours. This simplifying assumption has the advantage of making
analytic solutions to the Fokker-Plank equation possible.

\section{Analytical solutions}

\subsection{Solving for a fixed boundary}

As a first step towards understanding the formation of 
feeding-fronts, the response of an urchin population to a step-change
in the motility is considered. The boundary between the barren and the kelp regions is assumed to be fixed, with the macrophyte density being
greater than the critical density, $s_c$, for $x>0$ and less than $s_c$ for $x\leq 0$. It
follows from eq.~(\ref{eq:Dsc}) that the motility is $D = D_+$,
($x>0$) and $D = D_-$, ($x \leq 0$), where $D_\pm =
\lambda^2_\pm/2\Delta t$. If it is assumed that the urchin population
is initially uniformly distributed, then $u(x, 0) = u_\infty$, where
$u_\infty$ is a constant.

Away from the boundary between the two regions, the motility is
constant and eq.~(\ref{eq:fp}) reduces to a diffusion equation. If we
write $u(x, t) = u_+(x, t)$, ($x \geq 0$) and $u(x, t) = u_-(x, t)$,
($x < 0 $) then, for the derivatives on the right hand side of
eq.~(\ref{eq:fp}) to be continuous, we require that
\begin{equation}
\label{eq:continuity}
D_+u_+(0,t) = D_-u_-(0,t).
\end{equation}
We will look for a solution which has both $u_+(0,t)$ and $u_-(0,t)$
constant with time, and so will require that $\partial^2 (D u)/\partial x^2 |_{x=0} =
0$. Because the total urchin population is constant, any increase in the urchin density at positive $x$ must be matched by a decrease in density at negative $x$, 
\begin{equation}
\label{eq:areas}
\int_0^\infty (u_+-u_\infty) dx  = \int_{-\infty}^0
(u_\infty - u_-) dx .
\end{equation}  
The solution to a diffusion equation with a constant boundary is given
by the complementary error function, 
\begin{equation}
  \label{eq:erfc}
  \mathrm{erfc}(x) =  1 - {1 \over \sqrt{\pi}} \int_0^x e^{-\beta^2} d\beta,
\end{equation}
with $\beta$ being an integration constant. The solution
for the urchin population may be written as
\begin{equation}
  \label{eq:uprime}
  u_\pm(x, t) = u_\infty\left(1 \mp \gamma_\pm \mathrm{erfc}\left(|x|/2\sqrt{D_\pm t}\right)\right),
\end{equation}
where $\gamma_\pm$ are constants which must satisfy
\begin{equation}
  \label{eq:gamma}
  D_+\gamma_+ + D_-\gamma_- = D_+ - D_-
\end{equation}
in order to solve eq.~(\ref{eq:continuity}). For eq.~(\ref{eq:areas}) to hold,
\begin{equation}
  \label{eq:area}
  \gamma_+/\gamma_- = \sqrt{D_-/D_+}.
\end{equation}
With this ratio $\partial^2 Du/\partial x^2|_{x=0} = 0$, and the
Fokker-Planck equation is solved throughout the domain. From
eqs.~(\ref{eq:gamma}) and (\ref{eq:area}) it follows that
\begin{equation}
  \label{eq:gammapm}
  \gamma_\pm = {D_+ - D_- \over \sqrt{D_\pm}(\sqrt{D_-}+\sqrt{D_+})}.
\end{equation}

A plot of the solution is given in fig.~(\ref{fig:errorfunc}). The initially uniform urchin density develops a peak at the boundary between the two regions. There is an increased urchin density just inside the kelp, and a depleted region on the barren side of the boundary. The height of the peak is constant with time, but the width grows steadily. On the barren side of the peak there is a region where the sea-urchin density is less than the initial value.

\subsection{Solving for a moving boundary}

We now look for  traveling wave solutions of the Fokker-Planck
equation, representing a steadily moving urchin front. At this stage, the grazing of the urchins is not considered, it is simply assumed that the boundary between the two regions moves at a constant velocity $c$. The variable
$z = x - ct$ is introduced. The traveling solutions are functions of $z$ only, and
they satisfy the equation, derived from eq.~(\ref{eq:fp}),
\begin{equation}
  \label{eq:fpz}
  -c {du \over dz} = {d^2 Du \over dz^2},
\end{equation}
where $u = u(z)$ and $D=D(z)$. If the boundary between the grazed and ungrazed regions falls at $z=0$, the motility is
\begin{equation}
  \label{eq:Dz}
  D(z) = \left\{ 
    \begin{array}{cl}
    D_-, & z\leq 0 \\
    D_+, & z > 0 \\
    \end{array}
    \right. .
\end{equation}
By integrating eq.~(\ref{eq:fpz}) twice, an integral equation for the urchin density is obtained, 
\begin{equation}
 \label{eq:Du}
  D(z)u(z) = - c \int_{-\infty}^z (u(x) - u_\infty) dx + D_-u_\infty ,
\end{equation}
where the constant of integration, $u_\infty$, has been chosen so that $u(\pm \infty) = u_\infty$.

It is straightforward to verify that the solution to eq.~(\ref{eq:Du}) is given by the function
\begin{equation}
 \label{eq:usolution}
  {u(z)\over u_\infty} =  \left\{ 
  \begin{array}{ll}
   1, & z \leq 0 \\
   {D_- - D_+ \over D_+}e^{-cz/D_+} + 1, & z>0
   \end{array}
  \right. .
\end{equation}
If the motility is larger in the grazed region, $D_- > D_+$, then the traveling wave solution has the form of a feeding front, with a peak at the boundary between the regions. The maximum density within the feeding front occurs on the boundary, with a density $u_\infty D_-/D_+$. The urchin density is constant throughout the barren region, and decays exponentially towards the ungrazed side of the boundary, the front having a width of $D_+/c$. 

The feeding front can only propagate continually if there is a non-zero urchin density within the ungrazed region. Otherwise, the front will lose urchins as it travels and decay away.

\section{Introducing seaweed}
\begin{figure}
\label{fig:ghplot}
\centering
\includegraphics[width=7cm]{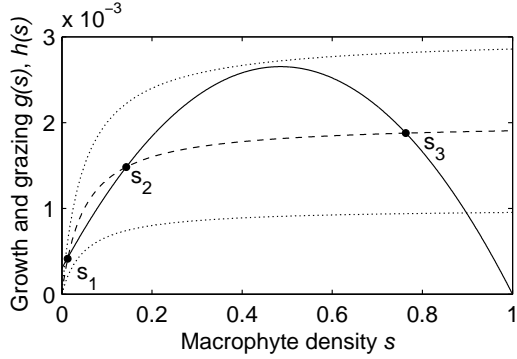}
\caption{Variation in macrophyte growth $g(s)$ (solid line)  and
  urchin grazing $h(s)$ (dashed line) as a function of the macrophyte
  density, $s$. The curves follow eqs.~(\ref{eq:gs}, \ref{eq:hs}),
  with the parameters $\mu_s = 0.01$ day$^{-1}$; $s_0 =
  0.03$ $s_{\max}$;
  $k_s = 0.05$ $s_{\max}$; $\alpha = 0.001$ $s_{\max}$ urchin$^{-1}$ day$^{-1}$. The dashed line is drawn for an
  urchin density of $u_\infty = 2$ urchin m$^{-2}$. The three intersection points of
  $g(s)$ and $h(s)$ are labeled by the macrophyte densities $s_1$,
  $s_2$ and $s_3$. The upper and lower dotted lines show the urchin
  grazing with the same parameters, but with urchin densities of
   $u_\infty = 3$ urchin m$^{-2}$ and  $u_\infty = 1$ urchin
  m$^{-2}$, respectively. With these densities there is only one
  solution of $ds/dz = 0$ (eq.~\ref{eq:dsdz}), and so there are no
  possible  traveling wave solutions that could represent an urchin feeding-front.}
\end{figure} 

Having identified a frontal solution to the urchin density when the boundary is moving steadily, the question is whether there are
traveling wave solutions to the coupled seaweed-urchin equations. The change in algal density is taken to occur through a combination
of growth and grazing,
\begin{equation}
  \label{eq:dsdt}
  {\partial s \over \partial t } = g(s) - h(s)u,
\end{equation}
where $g(s)$ describes the algal growth and $h(s)u$ is
the grazing rate of the urchins on the seaweed.  There is no explicit seaweed
dispersal included.  Recruitment from a wider
seaweed population is simply represented by a non-zero intercept of
$g(s)$.
For a traveling wave solution to exist, $s$ must be a function
of $z = x - ct$ only, so
\begin{equation}
  \label{eq:dsdz}
  {\partial s \over \partial z } = h(s) u/c    - g(s)/c.
\end{equation} 
At $z=\pm \infty$ the population must be in equilibrium, $h(s)u_\infty
= g(s)$, with $s(\infty) > s_c$ and $s(-\infty) \leq s_c$.  There must
be at least three real, positive solutions to 
\begin{equation}
  \label{eq:condition}
  \partial s/\partial z|_{u=u_\infty} =0,
\end{equation}
which we shall call $s_1$, $s_2$, and $s_3$ ($s_3>s_2>s_1$). The
solutions $s_1$ and $s_3$ are stable, and $s_2$ is unstable. In order
that $s<s_c$ for $z<0$ it is required that 
\begin{equation}
  \label{eq:scbound}
  s_2>s_c>s_1.
\end{equation}
If this does not
hold then no traveling wave solutions can be obtained. The propagation
speed can be obtained by requiring that $s = s_c$ at $z=0$, where the
urchin density, $u$,
in eq.~(\ref{eq:dsdz}) is obtained from eq.~(\ref{eq:usolution}).

As a plausible example, assume that macrophyte growth is logistic
\begin{equation}
  \label{eq:gs}
  g(s) = \mu_s (s + s_0)(1 - s/s_{\max}),
\end{equation} 
where $\mu_s$ (dimensions $[t^{-1}]$) is the growth-rate
$s_{\max}$(dimensions $[s]$) is the macrophyte carrying
capacity and the term $\mu_s s_0$ (dimensions $[s t^{-1}]$) represents
a background recruitment rate. With this growth function, the
macrophyte will grow to a density $s_{\max}$ in the absence of
urchins, and this growth to a maximal density will take a time of
order $\mu_s^{-1}$. 

An appropriate representation of grazing is the Holling
type II or Michaelis-Menten equation \citep{Holling59, Begon96}
\begin{equation}
  \label{eq:hs}
  h(s) = { \alpha s \over s + k_s},
\end{equation} 
where $\alpha$ (dimensions $[s u^{-1} t^{-1}]$) parameterizes the maximal grazing rate per urchin,
and $k_s$ (dimensions $[s]$) is the half-saturation constant for urchin grazing. At low
algal densities the grazing function decreases to zero,
representing the difficulty that urchins have in locating food when the macrophyte is sparse.

As an example, growth parameters relevant to the New Zealand
alga \emph{Ecklonia radiata} are used. This species grows to a mature
size within a year, and so an order-of-magnitude growth-rate is estimated to be     
$\mu_s = 0.01$ day$^{-1}$. The recruitment density $s_0$ will be site
specific, depending on the abundance of mature alga in the surrounding
area. It is simply assumed that $s_0$ is a small fraction of the
maximum density, $s_0 = 0.03 s_{\max}$.  An estimate of urchin
grazing rates may be obtained from the results of a small experiment carried out by Russell Cole (1993). A square
meter quadrat was loaded with urchins (\emph{Evechinus chloroticus}),
to a density of 60 m$^{-2}$, and the decrease in the abundance of the
alga \emph{E. radiata} was monitored. Even at this high urchin
density the decline in alga was slow, with a time-scale of $\sim$ 20
days. The maximum grazing rate is therefore $\alpha =
1/(20\times 60) = 0.001$ $s_{\max}$ urchin$^{-1}$ m$^{2}$ day$^{-1}$. The
algal density at which the urchin grazing is half of its maximum is
taken to be $k_s = 0.1 s_{\max}$. In the absence of any data on the
variation of urchin motility with algal concentration, it will simply
be assumed that the critical algal density is $s_c = k_s$. The growth
and grazing curves that result from these parameters are shown in
fig.~(\ref{fig:ghplot}), for three differing urchin densities. Detailed
experiment would be needed to verify both the functional form and the
parameterization of the growth and grazing functions. The intent here
is to illustrate the qualitative features of the urchin-macrophyte
system, rather than quantitative modeling of a specific case.

The existence of three solutions to
eq.~(\ref{eq:condition}) could be determined by directly solving this cubic
equation. While analytically tractable, the general solution will be
complicated. A more amenable estimate of when three real, positive solutions can
be found is readily obtained by graphical inspection of the growth and
grazing functions, $g(s)$ and $h(s)$. If the recruitment density $s_0$
is zero, then three solutions to eq.~(\ref{eq:condition}) will only be
found if the initial slope of the grazing function is larger than the
initial slope of the growth function. This will only hold if
\begin{equation}
  \label{eq:ugreater}
  u_\infty > {\mu_s k_s \over \alpha}.
\end{equation}
If $k_s << s_{\max}$,  then the maximal grazing rate  also needs to be less than
the maximal growth rate. This implies that 
\begin{equation}
  \label{eq:uless}
  u_\infty < {\mu_s s_{\max} \over 4 \alpha}.
\end{equation}
 Both of these
inequalities, (\ref{eq:ugreater}) and (\ref{eq:uless}), can only be
satisfied simultaneously if
\begin{equation}
  \label{eq:kcondition}
  k_s < s_{\max}/4.
\end{equation}
If the recruitment density $s_0$ is non-zero but small, $s_0 <<
s_{\max}$, then these conditions will still be relevant. For the
parameters used in fig.~(\ref{fig:ghplot}) the conditions given in
eqs.~(\ref{eq:ugreater}, \ref{eq:uless}) translate to the requirement
that 1 urchin m$^{-2}$ $< u_{\infty} <$ 2.5 urchin m$^{-2}$. These are not exact bounds, but they provide a useful estimate of the range
over which three solutions to eq.~(\ref{eq:dsdz}) can be found. 

For a feeding-front solution to exist it is also necessary that the
transition from high to low urchin motility occurs at a macrophyte
density, $s_c$, which is between $s_1$ and $s_2$
(eq.~\ref{eq:scbound}). In the case presented in
fig.~(\ref{fig:ghplot}), this would be satisfied by $s_c = 0.1
s_{\max}$. The range of initial urchin densities over
which a feeding front solution develops is small, with a factor of less than 3 between a density that leads to macrophyte beds and a  density
that results in urchin barrens.  

\section{Numerical simulations}
\label{sec:numsim}

\begin{figure}
	\label{fig:peakcontour}
	\centering
	\includegraphics[width=7cm]{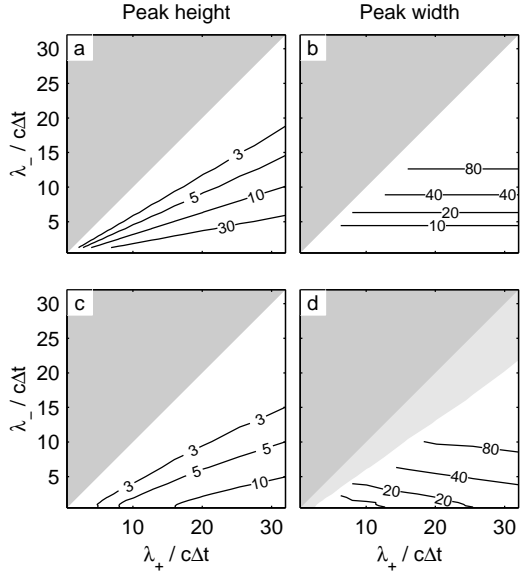}
	\caption{Comparison between the theoretical form of a traveling wave
	(eq.~\ref{eq:usolution}) and numerical simulations, as described in
	section \ref{subsec:traveling_wave}. (a) The theoretical peak height, $D_-/D_+$ (b) The
	theoretical width, $D_+/c$. (c) The maximum height
	of the peak from the simulations (d) The width of the peak. The half-shading masks the region where the peak height is too small to allow a width to be reliably calculated.}
\end{figure}

\subsection{The traveling wave}
\label{subsec:traveling_wave}
For comparison with the analytic solutions numerical simulations are carried out.
The first set of simulations aims to check the validity of the traveling wave 
solution, eq.~(\ref{eq:usolution}). A one dimensional model is built, which 
begins with uniformly distributed urchins, each urchin having a real-valued 
position. The boundary between the low and high motility regions begins 
at $x = 1200$ m and moves towards the right at a velocity $c = 1$ m day$^{-1}$.
At each timestep, for each urchin, a random number $\eta$ is generated from a normal
distribution with zero mean and variance. If 
the urchin is to the right of the boundary  it is moved by $\lambda_+ \eta$. 
Otherwise, the urchin is moved by  $\lambda_- \eta$. These rules capture 
the assumptions which led to the derivation of eq.~(\ref{eq:usolution}). A
window 800 m wide is maintained around the boundary, with a border 150 m 
wide beyond that. Urchins are added or removed from the simulation to hold 
the density constant within the two border regions. Any urchins which move 
beyond the border are removed.  The simulation starts with a uniform density
of 50 urchins m$^{-1}$. It is run for 2000 timesteps, with data from the final
200 timesteps being grouped into 1 m long bins and averaged.  The whole 
simulation is repeated for a range of $\lambda_+$ and $\lambda_-$ ($\lambda_- > \lambda_+$).  A comparison 
of the theoretical and the numerical peak widths and heights are shown in 
fig.~(\ref{fig:peakcontour}). There is good agreement between the two approaches, 
confirming that these simple assumptions can lead to a propagating peak in urchin 
density.

\subsection{Two dimensional simulations with macrophyte}

\begin{figure}
	\label{fig:seaweed}
	\centering
	\includegraphics[width=7cm]{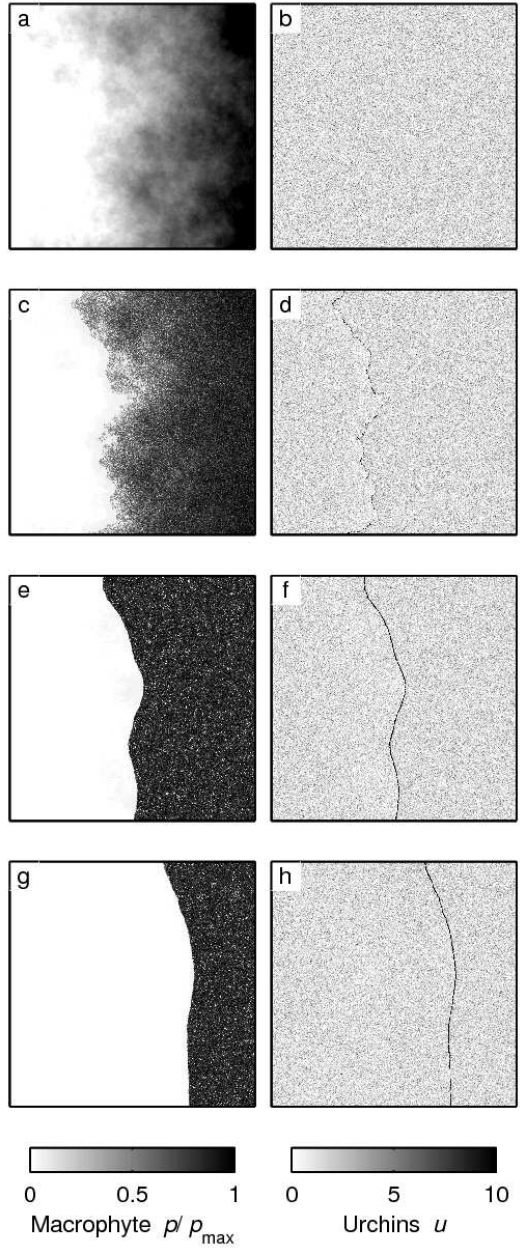}
	\caption{Simulation of seaweed (a, c, e, g) and urchins (b, d, f, h), showing the 
	formation of a feeding front. The pictures are made at 0 (a, b), 600 (c, d), 
	3000 (e, f) and 6000 (g, h) model days.}

\end{figure}

Finally, a simulation is run to check the stability of the
feeding fronts in a two-dimensional setting, with macrophyte. A numerical domain
is used which represents a 500 m $\times$ 500 m square, divided into 1 m$^2$ cells. 
Each cell has a seaweed density, $s$, with the density going from $s = 0$ on the left 
hand side of the domain to $s = s_{\max}$ on the right hand side. The seaweed distribution 
has some initial variability, introduced by adding a random function to the linear gradient 
(fig.~\ref{fig:seaweed}a). The random function has greater variability at longer length 
scales, with a Fourier transform that decays as $f^{-3/4}$, where $f$ is the wavenumber. 
This is done to introduce noise into the model, capturing in some way the natural environmental
variability. Urchins are then added, uniformly distributed through the whole domain, 
and with an average density of 1.5 urchin m$^{-2}$. At each timestep the seaweed within each cell changes
according to eq.~(\ref{eq:dsdt}). A simple finite-difference approximation is used, and 
the seaweed density is always kept above zero. The urchin density is calculated from the
number of urchins within each 1 m$^2$ cell, and the seaweed is grazed accordingly. The 
urchins are then moved by a random amount, with the size of the step, $\lambda$, depending 
on whether the seaweed density exceeds the threshold. Any urchins moving outside the
domain are reflected back into it, so the total number of urchins within the domain is constant.

The results are shown in fig.~(\ref{fig:seaweed}). In the simulations shown, the following 
parameters have been used, $\lambda_+ = 0.05$ m day$^{-1}$; $\lambda_- = 1$ m day$^{-1}$;  
$k_s = 0.05 s_{\max}$; $s_0 = 0.01 s_{\max}$; $\mu = 0.01$ day$^{-1}$; $\alpha = 0.001
s_{\max}$ urchin$^{-1}$ m$^{2}$; and $s_{c} = k_s$, similar to the parameters in fig.~(\ref{fig:ghplot}). 
The simulations are run for 10,000 model days, with the figure showing 
the seaweed density and the urchin density at 0, 600, 3000 and 6000 model days.

From the start of the simulation the seaweed density becomes increasingly polarized, 
with areas of urchin barren, and areas of close to maximum density. A feeding front 
develops along the boundary between the regions, and the boundary slowly propagates 
towards the ungrazed region. The front appears stable, becoming smoother with time.

\section{Discussion}

The simple assumptions of differential urchin movement in response to seaweed density 
lead to the formation and propagation of an urchin feeding front, in qualitative agreement 
with observations. No social behavior needs to be assumed to explain the persistence of 
the front, and the motion of each urchin can be random. The system provides an excellent example of how simple individual processes can lead to spatial pattern. The development of the fronts shows the importance of correctly representing 
movement. Diffusion approximations, based on Fickian diffusion, are often used to represent 
animal dispersal \citep{Okubo80}. Because Fickian diffusion will always lead to the density of a population
decreasing (at least in the absence of any reproduction or migration) it is unable to generate sharp fronts.  A simple change to the representation of dispersal, from Fickian to Fokker-Planck, leads to a model that captures the qualitative features of the system. The focus of the analysis has been on demonstrating that the system can develop a stable propagating front. This simple model may also be used to explore the dynamics of more transient phenomena, such as the effect of a localised recruitment of urchins, or how a patchy mosaic of barrens and macrophyte habitat can be maintained. With the two states that are stable to small perturbations and the transitional wave transforming one to the other, the urchin-macrophyte system has many of the features of excitable media \citep{Murray93}. There is ample scope for further exploration of this analogy.

With the movement rates used here, the propagation speed of the front is very slow. 
In the two dimensional simulation, the front moved at a speed of 10 m per model year. This is on a similar order to propagation speeds of 2.5 m month$^{-1}$ reported from field observations \citep{Gagnon04}.
In contrast the aggregation of \emph{Lytechinus variegatus} in Florida Bay was reported to
move at 6 m day$^{-1}$. The propagation rate will be strongly dependent on the details of 
the urchin grazing. It is likely that the assumption of asociality, or of urchin independence, breaks down at the
high densities encountered in the front. Because of the very narrow spatial extent of the frontal region, the urchin behavior at high densities will effect the outcome of the model. To produce a quantitatively accurate model would 
require more detailed observations. Studies which focus on the movements of individual urchins \citep{Lauzonguay06} are likely to generate the data required to build a better representation of the frontal dynamics. For example the inclusion of a traffic-jam effect, where the movement rate of the urchins decreases as the density increases \citep{Lauzonguay06}, will result in an increased urchin density within the front.

As discussed in the introduction, there are other processes that are known to influence urchin behaviour which could be represented within a model of this nature. Unfortunately, the effects of many of these factors have only been measured in a few isolated experiments, and there is insufficient published data to include them. The model developed here is in many ways a null model. It is hoped that it will inspire experimentalists to collect the individual based data which is needed to understand the full detail of how urchin feeding fronts are formed and maintained.

\section{Acknowledgements}
I am grateful to Andrew Visser for discussions on random walks; to Russell Cole for discussions on urchin biology and to Alison MacDiarmid and Alistair Dunn for their support and encouragement. This project was funded by the New Zealand Foundation for Research in Science and Technology.

\bibliography{kk}

\begin{thebibliography}{35}
\expandafter\ifx\csname natexlab\endcsname\relax\def\natexlab#1{#1}\fi
\expandafter\ifx\csname url\endcsname\relax
  \def\url#1{\texttt{#1}}\fi
\expandafter\ifx\csname urlprefix\endcsname\relax\def\urlprefix{URL }\fi

\bibitem[{Alcoverro(2002)}]{Alcoverro02}
Alcoverro, 2002. Effects of sea urchin grazing on seagrass
  (\emph{Thalassodendron ciliatum}) beds of a {K}enyan lagoon. Mar. Ecol. Prog.
  Ser. 226, 255--263.

\bibitem[{Andrew and Stocker(1986)}]{Andrew86}
Andrew, N.~L., Stocker, L.~J., 1986. Dispersion and phagokinesis in the
  echinoid \emph{Evechinus chloroticus} ({V}al.). J. Exp. Mar. Biol. Ecol. 100,
  11--23.

\bibitem[{Begon et~al.(1996)Begon, Harper, and Townsend}]{Begon96}
Begon, M., Harper, J.~L., Townsend, C.~R., 1996. Ecology: Individuals,
  populations and communities, 3rd Edition. {B}lackwell {S}cience, Cambridge,
  {MA}.

\bibitem[{Bernstein et~al.(1981)Bernstein, Williams, and Mann}]{Bernstein81}
Bernstein, B.~B., Williams, B.~E., Mann, K.~H., 1981. The role of behavioral
  responses to predators in modifying urchins' (\emph{Strongylocentrotus
  droebachiensis}) destructive grazing and seasonal foraging patterns. Mar.
  Biol. 63, 39--49.

\bibitem[{Dance(1987)}]{Dance87}
Dance, C., 1987. Patterns of activity of the sea urchin \emph{Paracentrotus
  lividus} in the {B}ay of {P}ort-{C}ros ({V}ar, {F}rance, {M}editerranean).
  Mar. Ecol. 8, 131--142.

\bibitem[{Dare(1982)}]{Dare82}
Dare, P.~J., 1982. Notes on the swarming behaviour and population density of
  \emph{Asterias rubens} {L}. ({E}chindermata: {A}steroidea) feeding on the
  mussel, \emph{Mytilus edulis} {L}. J. Const. int. Explor. Mer. 40, 112--118.

\bibitem[{Dayton et~al.(1984)Dayton, Currie, Gerrodette, Keller, Rosenthal, and
  {V}en {T}resca}]{Dayton84}
Dayton, P.~K., Currie, V., Gerrodette, T., Keller, B.~D., Rosenthal, R., {V}en
  {T}resca, D., 1984. Patch dynamics and the stability of some {C}alifornia
  kelp communities. Ecol. Monogr. 54, 253--289.

\bibitem[{Dean et~al.(1984)Dean, Schroeter, and Dixon}]{Dean84}
Dean, T.~A., Schroeter, S.~C., Dixon, J.~D., 1984. Effects of grazing by two
  species of sea urchins (\emph{Strongylocentrotus franciscanus} and
  \emph{Lytechinus anamesus}) on recruitment and survival of two species of
  kelp (\emph{Macrocystis pyrifera} and \emph{Pterygophora californica}). Mar.
  Biol. 78, 301--313.

\bibitem[{Dix(1970)}]{Dix70}
Dix, T.~G., 1970. Biology of \emph{Evechinus chloroticus} ({E}chinoidea:
  {E}chinometridae) from different localities. N. Z. J. Mar. Freshwat. Res. 4,
  267--277.

\bibitem[{Duggan and Miller(2001)}]{Duggan01}
Duggan, R.~E., Miller, R.~J., 2001. External and internal tags for the green
  sea urchin. J. Exp. Mar. Biol. Ecol. 258, 115--122.

\bibitem[{Dumont et~al.(2004)Dumont, Himmelman, and Russel}]{Dumont04}
Dumont, C., Himmelman, J.~H., Russel, M.~P., 2004. Size-specific movement of
  green sea urchins \emph{Strongylocetrotus droebachiensis} on urchin barrens
  in eastern {C}anada. Mar. Ecol. Prog. Ser. 276, 93--101.

\bibitem[{Dumont et~al.(2006)Dumont, Himmelman, and Russel}]{Dumont06}
Dumont, C., Himmelman, J.~H., Russel, M.~P., 2006. Daily movement of the sea
  urchin \emph{Strongylocetrotus droebachiensis} in different subtidal habitats
  in eastern {C}anada. Mar. Ecol. Prog. Ser. 317, 87--99.

\bibitem[{Gagnon et~al.(2004)Gagnon, Himmelman, and Johnson}]{Gagnon04}
Gagnon, P., Himmelman, J.~H., Johnson, L.~E., 2004. Temporal variation in
  community interfaces: kelp-bed boundary dynamics adjacent to persistent
  urchin barrens. Mar. Biol. 144, 1191--1203.

\bibitem[{Hagen(1995)}]{Hagen95}
Hagen, N.~T., 1995. Recurrent destructive grazing of successionally immature
  kelp forests by green sea urchins in {V}estfjorden, {N}orthen {N}orway. Mar.
  Ecol. Prog. Ser. 123, 95--106.

\bibitem[{Hart and Chia(1990)}]{Hart90}
Hart, L.~J., Chia, F.-S., 1990. Effect of food supply and body size on the
  foraging behavior of the burrowing sea urchin \emph{Echinometra mathaei} (de
  {B}lainville). J. Exp. Mar. Biol. Ecol. 135, 99--108.

\bibitem[{Holling(1959)}]{Holling59}
Holling, C.~S., 1959. Some characteristics of simple types of predation and
  parasitism. Canadian Entomologist 91, 385--398.

\bibitem[{Kareiva and Odell(1987)}]{Kareiva87}
Kareiva, P., Odell, G., 1987. Swarms of predators exhibit ``preytaxis'' if
  individual predators use area-restricted search. Am. Nat. 130, 233--270.

\bibitem[{Kawamata(1998)}]{Kawamata98}
Kawamata, S., 1998. Effect of wave-induced oscillatory flow on grazing by a
  subtidal sea urchin \emph{Strongylocentrotus nudus} ({A}. {A}gassiz). J. Exp.
  Mar. Biol. Ecol. 224, 31--48.

\bibitem[{Keller and Segel(1971)}]{Keller71b}
Keller, E.~F., Segel, L.~A., 1971. Traveling bands of chemotactic bacteria: a
  theoretical analysis. J. theor. Biol. 30, 235--248.

\bibitem[{Klinger and Lawrence(1985)}]{Klinger85}
Klinger, T.~S., Lawrence, J.~M., 1985. Distance perception of food and the
  effect of food quantity on feeding behavior of \emph{Lytechinus variegatus}
  ({L}amarck) ({E}chinodermata: {E}chinoidea). Mar. Behav. Physiol. 11,
  327--344.

\bibitem[{Konar and Estes(2001)}]{Konar01}
Konar, B., Estes, J.~A., 2001. Seasonal changes in subarctic sea urchin
  populations from different habitats. Polar Biology 24, 754--763.

\bibitem[{Konar and Estes(2003)}]{Konar03}
Konar, B., Estes, J.~A., 2003. The stability of boundary regions between kelp
  beds and deforested areas. Ecology 84, 174--185.

\bibitem[{Laur et~al.(1986)Laur, Ebeling, and Reed}]{Laur86}
Laur, D.~R., Ebeling, A.~W., Reed, D.~C., 1986. Experimental evaluations of
  substrate types as barriers to sea urchin ({S}trongylocentrotus spp.)
  movement. Mar. Biol. 93, 209--215.

\bibitem[{Lauzon-Guay et~al.(2006)Lauzon-Guay, Scheibling, and
  Barbareau}]{Lauzonguay06}
Lauzon-Guay, J.-S., Scheibling, R.~E., Barbareau, M.~A., 2006. Movement
  patterns in the green sea urchin, \emph{Strongylocentrotus droebachiensis}.
  J. Mar. Biol. Ass. U.K. 86, 167--174.

\bibitem[{Maci\'a and Lirman(1999)}]{Macia99}
Maci\'a, S., Lirman, D., 1999. Destruction of {F}lorida {B}ay seagrasses by a
  grazing front of sea urchins. Bull. Mar. Sci. 65, 593--601.

\bibitem[{Mattison et~al.(1977)Mattison, Trent, Shanks, Akin, and
  Pearse}]{Mattison77}
Mattison, J.~E., Trent, J.~D., Shanks, A.~L., Akin, T.~B., Pearse, J.~S., 1977.
  Movement and feeding activity of red sea urchins (\emph{Strongylocentrotus
  francsicanus}) adjaceant to a kelp forest. Mar. Biol. 39, 25--30.

\bibitem[{Murray(1993)}]{Murray93}
Murray, J.~D., 1993. Mathematical biology, 2nd Edition. Vol.~19 of
  Biomathematics. Springer-Verlag, Berlin.

\bibitem[{Okubo(1980)}]{Okubo80}
Okubo, A., 1980. Diffusion and ecological problems: mathematical models.
  Vol.~10 of Biomathematics. Springer-Verlag, Berlin.

\bibitem[{Pisut(2002)}]{Pisut02}
Pisut, D.~P., 2002. The distance chemosensory foraging behavior of the sea
  urchin \emph{Lytechinus variegatus}. Master's thesis, Georgia {I}nstitute of
  {T}echnology, {A}tlanta, {G}eorgia.

\bibitem[{Scheibling et~al.(1999)Scheibling, Hennigar, and
  Balch}]{Scheibling99}
Scheibling, R.~E., Hennigar, A.~W., Balch, T., 1999. Destructive grazing,
  epiphytism, and disease: the dynamics of sea urchin - kelp interactions in
  {N}ova {S}cotia. Can. J. Fish. Aquat. Sci./J. can. sci. halieut. aquat. 56,
  2300--2314.

\bibitem[{Stoner(1989)}]{Stoner89}
Stoner, A.~W., 1989. Winter mass migration of juvenile queen conch
  \emph{Strombus gigas} and their influence on the benthic environment. Mar.
  Ecol. Prog. Ser. 56, 99--104.

\bibitem[{Stoner and Lally(1994)}]{Stoner94}
Stoner, A.~W., Lally, J., 1994. High-density aggregation in queen conch
  \emph{Strombus gigas}: formation, patterns, and ecological significance. Mar.
  Ecol. Prog. Ser. 106, 73--84.

\bibitem[{Turchin(1998)}]{Turchin98}
Turchin, P., 1998. Quantitative analysis of movement. Sinauer, Sunderland.

\bibitem[{Vadas et~al.(1986)Vadas, Elner, Garwood, and Babb}]{Vadas86}
Vadas, R.~L., Elner, R.~W., Garwood, P.~E., Babb, I.~G., 1986. Experimental
  evaluation of aggregation behavior in the sea urchin \emph{Strongylocentrotus
  droebachiensis}. Mar. Biol. 90, 433--448.

\bibitem[{Wright et~al.(2005)Wright, Dworjanyn, Rogers, Steinberg, Williamson,
  and Poore}]{Wright05}
Wright, J.~T., Dworjanyn, S.~A., Rogers, C.~N., Steinberg, P.~D., Williamson,
  J.~E., Poore, A. G.~B., 2005. Density-dependent sea urchin grazing :
  differential removal of species, changes in community composition and
  alternative community states. Mar. Ecol. 198, 143--156.

\end{thebibliography}

\end{document}